\documentclass[showpacs,preprintnumbers,amsmath,amssymb]{revtex4}
\usepackage{graphicx}
\usepackage{dcolumn}
\usepackage{bm}

\begin{document}
\title{Statistical properties of Cherenkov and quasi-Cherenkov superradiance}

\author{S.V.~Anishchenko}
\email{sanishchenko@mail.ru}
\affiliation{Research Institute for Nuclear Problems\\
Bobruiskaya str., 11, 220030, Minsk, Belarus.}%

\author{V.G.~Baryshevsky}
 \email{bar@inp.bsu.by}
 \affiliation{Research Institute for Nuclear Problems\\
Bobruiskaya str., 11, 220030, Minsk, Belarus.}%

\begin{abstract}
We consider the effects of shot noise and particle energy spread
on statistical properties of Cherenkov and quasi-Cherenkov superradiance emitted by a
relativistic electron beam.
 In the absence of energy spread, we have found the root-mean-square
 deviations of both peak radiated
power and instability growth time as a function of the number of
particles. It is shown that energy spread
can lead to a sharp drop in the radiated power of Cherenkov and quasi-Cherenkov superradiance at high currents.
\end{abstract}


\maketitle

\section{Introduction}
The numerous processes occurs in the electron bunch moving in the medium with the refractive index more than unity \cite{Ginzburg1996,McNeil1999,Wiggins2000,Ginzburg2002,Elchaninov2003,Ginzburg2013,AnishchenkoBaryshevsky2015}. Cherenkov instability is one of them.
As was demonstrated in \cite{Ginzburg1996,McNeil1999,Wiggins2000,Ginzburg2002}, the instability  evolves in the
electron bunch even in the
absence of an external action.  At
the initial stage, the instability is accompanied  by particle bunching and exponential growth of the radiated power. This exponential growth
is then stopped  due to nonlinearity,  and the pulse of
Cherenkov superradiance is formed.

The influence  of different parameters on the peak power of Cherenkov superradiance  was
studied in detail in \cite{Elchaninov2003}. The authors of \cite{Elchaninov2003} also assumed
 that the particles  having the same initial energy undergo phase
 pre-modulation at the radiation frequency.
 As a result of phase pre-modulation, the
 electromagnetic oscillations start  with spontaneous coherent
 emission
 from all particles rather than with spontaneous emission  from
 individual electrons of a bunch.

However, it seems interesting to know the behavior of Cherenkov superradiance without pre-modulation, in
which case the  generation starts as  spontaneous emission of
electromagnetic waves from individual particles.
To give a correct description of Cherenkov superradiance  in this
case, we need  to consider statistical fluctuations  due to shot
noise  and energy spread of electrons inherent in particle
ensembles \cite{Roberson1989,Tran1990,Penman1992,Fawley2002}.
It is well-known
\cite{Bonifacio1994,Saldin1998,Prazeres1997,Andruszkow2000} that
because of shot noise in
 single-pass FELs,  the radiated power and the instability growth time
become stochastic quantities whose
root-mean-square deviations have the same order of magnitude as
their average values.

This paper studies the statistical properties of Cherenkov and quasi-Cherenkov superradiance
without phase pre-modulation when the emission of electromagnetic
waves begins with spontaneous emission from individual particles.
The peak radiated power and the instability growth time are taken as
stochastic quantities for statistical analysis. The peak power is the main output
characteristic  of short-pulse sources of electromagnetic
radiation and the instability growth time is the parameter defining the
minimum particle passage time in the generator that is necessary for
the superradiant instability to evolve.

The paper's outline is as follows. First, we derive a system of
equations describing the interaction of charged particles with the radiation field in the medium.
Further comes a detailed consideration of statistical properties
of Cherenkov \cite{McNeil1999,Wiggins2000} and quasi-Cherenkov \cite{AnishchenkoBaryshevsky2015} superradiance in the presence of shot noise alone and
then the energy spread of electrons is added. The energy spread of
electrons will appear to be an important factor limiting the peak
radiated power of Cherenkov superradiance.

\section{Interaction between charged particles and the electromagnetic field}
Let us consider the electron bunch of length $L$. The bunch is directed by a strong longitudinal magnetic field. As a result, there is no transverse displacement of particles. The effective interaction between the bunch and the electromagnetic wave is provided by Cherenkov synchronism condition: the electron velocity $v_0$ is close to the phase velocity $v_{ph}$.

Within the beam region, the radiation field is assumed to have a form
\begin{equation}
\label{bwo1}
E=\text{Re}(E_0(x,t)e^{i\Omega(t-x/v_0)}), 
\end{equation}
where the frequency $\Omega$ satisfies the equation
\begin{equation}
\label{bwo2}
v_{ph}(\Omega)=v_0,
\end{equation}
and the slowly varying complex amplitude of the wave $E_0$ satisfies the conditions:
\begin{equation}
\label{bwo3}
\begin{split}
&\Big|\frac{1}{\Omega E_0}\frac{\partial E_0}{\partial t}\Big|\ll1,\\
&\Big|\frac{v_0}{\Omega E_0}\frac{\partial E_0}{\partial x}\Big|\ll1.\\
\end{split}
\end{equation}
In this case, the excitation equation for $E_0$ can be written as follows \cite{Ginzburg1978}
\begin{equation}
\label{bwo4}
\frac{1}{v_{gr}}\frac{\partial E_0}{\partial t}+\frac{\partial E_0}{\partial x}=-\frac{\beta^2K}{2}I_0,
\end{equation}
where $v_{gr}$ is the group velocity, $K$ is the coupling impedance, $\beta=\Omega/v_0$, and
\begin{equation}
\label{bwo5}
\begin{split}
&I_0=\frac{\beta}{\pi}\int_{x-\pi/\beta}^{x+\pi/\beta}I(x,t)e^{-i(\Omega t-\beta x)}dx\\
&=\frac{\beta q_e}{\pi}\sum_\alpha v_\alpha e^{-i(\Omega t-\beta x_\alpha)}\\
\end{split}
\end{equation}
is the slowly varying complex amplitude of the current $I$.

Now, let us turn to the analysis of the electron motion. For this purpose, we assume that all charged particles located within the interval $[x-\pi/\beta,x+\pi/\beta]$ are affected by the same force determined by the amplitude $E_0$ at point $x$. As a result, the equations of particle motion has the following form:
\begin{equation}
\label{bwo6}
\begin{split}
&\frac{d\gamma_\alpha}{dt}=\frac{q_ev_0}{mc^2}\text{Re}E_0e^{i\theta_\alpha},\\
&\frac{d\theta_\alpha}{dt}=\Omega\Big(\frac{v_{0\alpha}}{v(\gamma_\alpha)}-1\Big)
\approx\frac{\Omega}{2}\Big(\frac{1}{\gamma_\alpha^2}-\frac{1}{\gamma_{0\alpha}^2}\Big).\\
\end{split}
\end{equation}
Here, $q_e$ and $m$ are the charge and mass of the electron repectively, $\gamma_\alpha=1/\sqrt{1-v_\alpha^2/c^2}$ is the Lorenz factor, and $\theta_\alpha$ is the paticle phase in the electromagnetic wave.

Let us introduce the average current density:
\begin{equation}
\label{bwo7}
I_{av}=\frac{\beta}{2\pi}\sum_\alpha q_ev_\alpha\approx\frac{\beta}{2\pi}N_\lambda q_ev_{0},
\end{equation}
where $N_\lambda$ is the number of particles in the interval $[x-\pi/\beta,x+\pi/\beta]$.

Using \eqref{bwo7} and coordinate transformation
\begin{equation}
\label{bwo8}
x=\tilde x+v_0t,
\end{equation}
we rewrite equation \eqref{bwo4} as follows
\begin{equation}
\label{bwo9}
\frac{1}{v_{gr}}\frac{\partial E_0}{\partial t}+\frac{v_{gr}-v_0}{v_{gr}}\frac{\partial E_0}{\partial \tilde x}=-\frac{\beta^2K}{N_\lambda}\sum_\alpha e^{-i\theta_\alpha}.
\end{equation} 

For $v_0>v_{gr}$ and $\gamma_0\gg1$, the substitution of dimensionless quantities 
\begin{equation}
\label{bwo10}
\begin{split}
&\tau=C\beta t\sqrt[3]{v_0^2v_{gr}},\\
&C=\frac{eI_{av}K}{2\gamma_0^3mc^2},\\
&z=C\beta \tilde x\frac{\sqrt[3]{v_0^2v_{gr}}}{v_0-v_{gr}},\\
&F=\frac{eE_0}{\gamma_0^3mc^2\beta C^2}\frac{\sqrt[3]{v_0^2v_{gr}}}{v_{gr}},\\
&\nu=2C\gamma_0^2\sqrt[3]{\frac{v_{gr}}{v_0}},\\
&\xi=C\beta L\frac{\sqrt[3]{v_0^2v_{gr}}}{v_0-v_{gr}}\\
\end{split}
\end{equation} 
into  \eqref{bwo6} and \eqref{bwo9} yields to the set of equations
\begin{equation}
\label{bwo11}
\begin{split}
&\frac{\partial F}{\partial \tau}-\frac{\partial F}{\partial z}=-\frac{2}{N_\lambda}\sum_\alpha e^{-i\theta_\alpha},\\
&\frac{d^2\theta_\alpha}{d\tau^2}=-\Big(1+\nu\frac{d\theta_\alpha}{d\tau}\Big)^{3/2}\text{Re}(Fe^{i\theta_\alpha}),\\
\end{split}
\end{equation} 
which should be supplemented with boundary and initial conditions
\begin{equation}
\label{bwo12}
\begin{split}
&\dot\theta_\alpha=0,\\
&F(\xi,\tau)=0,\\
&\theta_\alpha=2\pi r_\alpha,
\end{split}
\end{equation} 
where $r_\alpha$ are random variables uniformly distributed in the interval $[0;1)$.

One of the main parameters of short-pulse sources is the convertion ratio equal to the peak radiated power to the electron flow power ratio. In dimensionless units, $\eta$ is given by the expression
\begin{equation}
\label{bwo13}
\eta=\frac{v_{gr}}{v_{0}}\frac{\nu|F_{peak}|^2}{8},
\end{equation} 
where $F_{peak}$ is the peak value of the dimensionless amplitude $F$.

Further, we shall use the reduced conversion ratio
\begin{equation}
\label{bwo14}
P_0=\frac{\nu|F_{peak}|^2}{8}\Big|_{z=0}
\end{equation} 
 instead of $\eta$.
The quantity $P_0$ differs from $\eta$ by the numerical factor $\frac{v_{gr}}{v_{0}}$.

\section{Shot noise}

As follows from  \eqref{bwo11}, the behavior of charged particles in the absence of the energy spread
is determined by three controlling  parameters: the bunch length $\xi$, the nonlinearity parameter $\nu$, and the number of particles $N_e$.
Hence, to explore the statistical properties of Cherenkov superradiance, we need to solve the set of equations \eqref{bwo11}
for various values of the controlling parameters. Because  the
 initial phases $\theta_\alpha(0)$ are randomly distributed,   the numerical
experiment with each triple of values of $\xi$, $\nu$, and $N_e$ must
be repeated many times. This procedure will give information about
statistical characteristics of Cherenkov superradiance, the most
important of which are the reduced conversion ratio $P_0$, the instability growth time $T_0$,  and
their relative root-mean-square deviations $\delta_P$ and
$\delta_T$.

\begin{figure}[ht]
\begin{center}
 \resizebox{65mm}{!}{\includegraphics{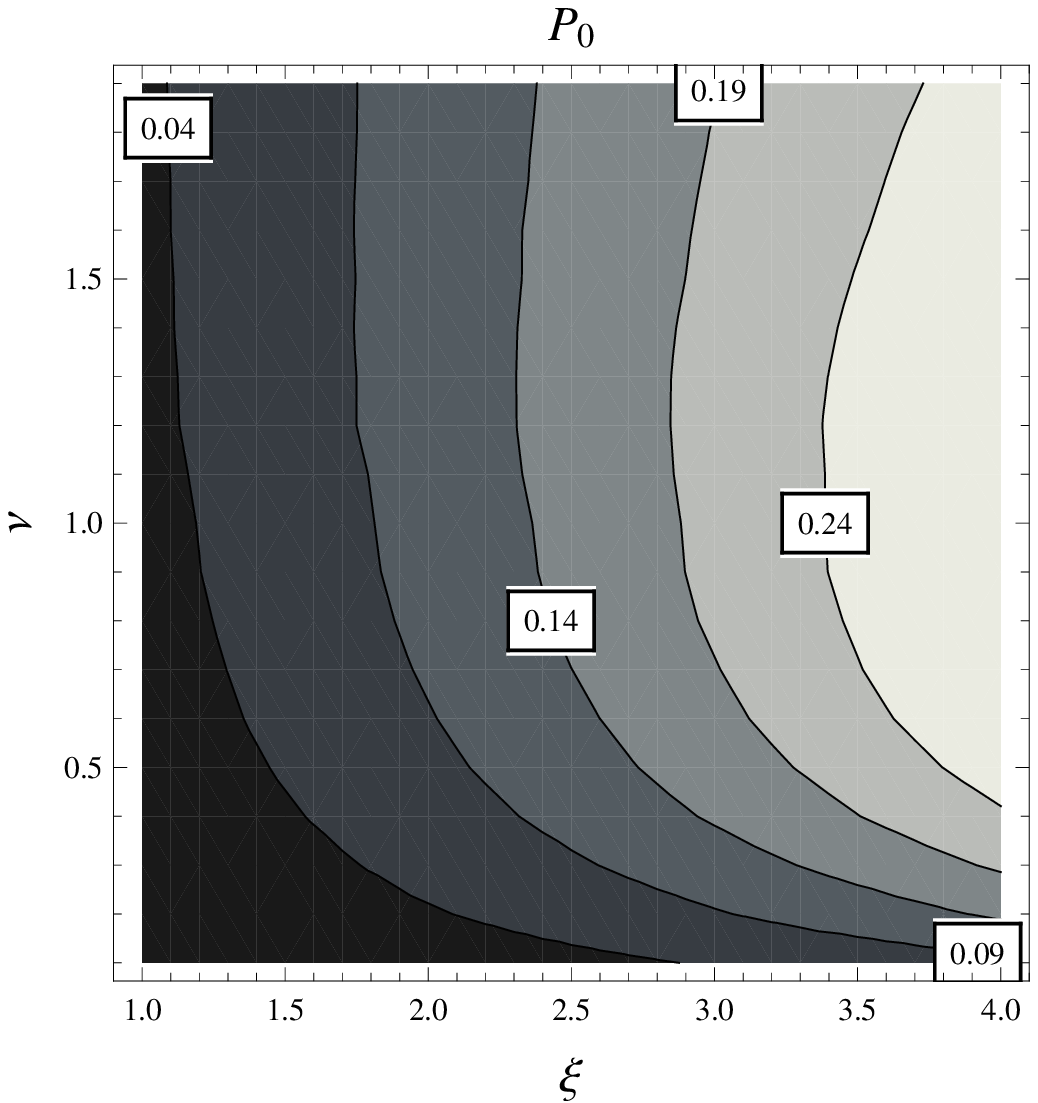}} \resizebox{65mm}{!}{\includegraphics{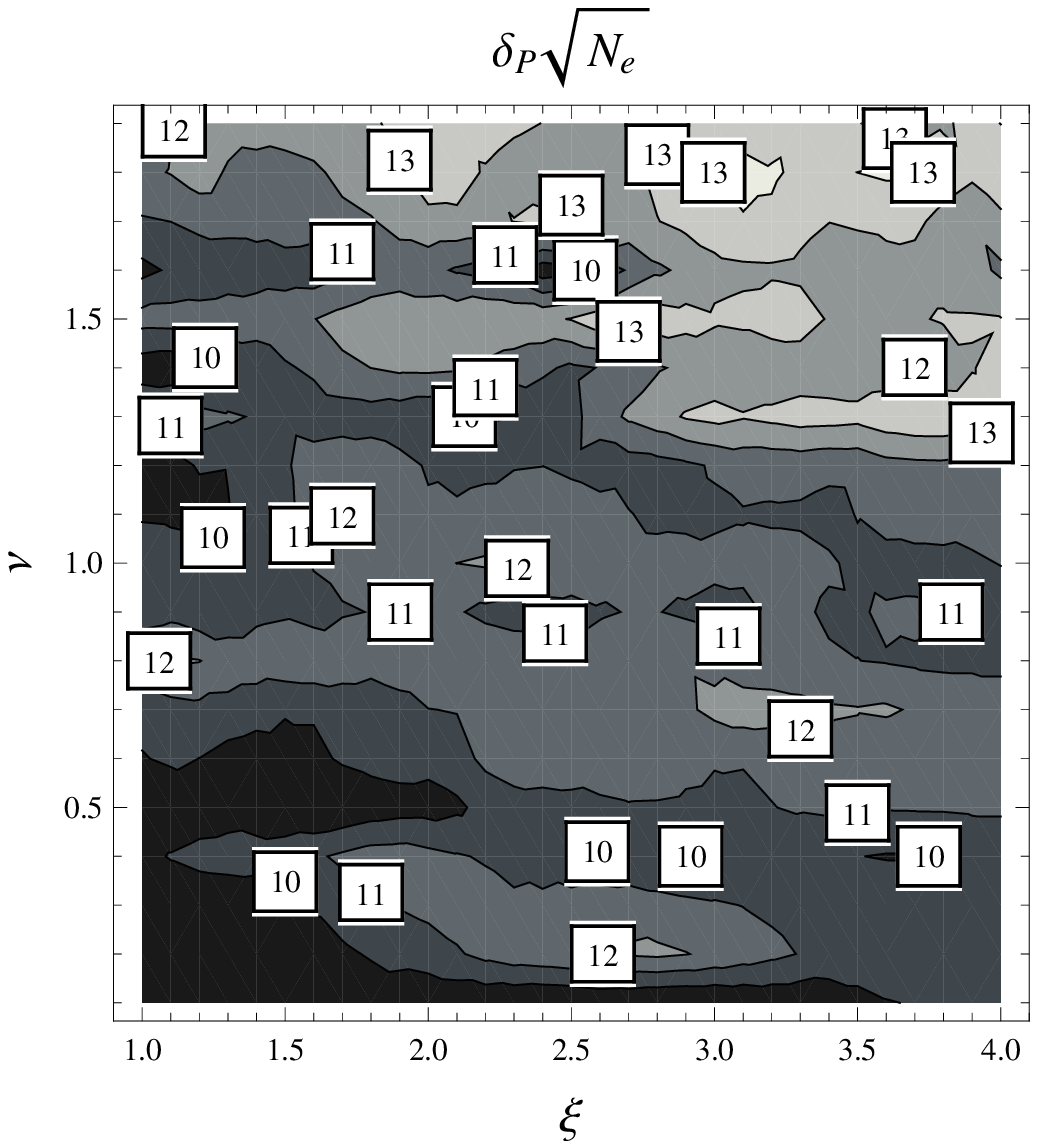}}\\
\end{center}
\caption{Reduced conversion ratio and its root-mean-square
deviation.} \label{Fig.1}
\end{figure}

\begin{figure}[ht]
\begin{center}
  \resizebox{65mm}{!}{\includegraphics{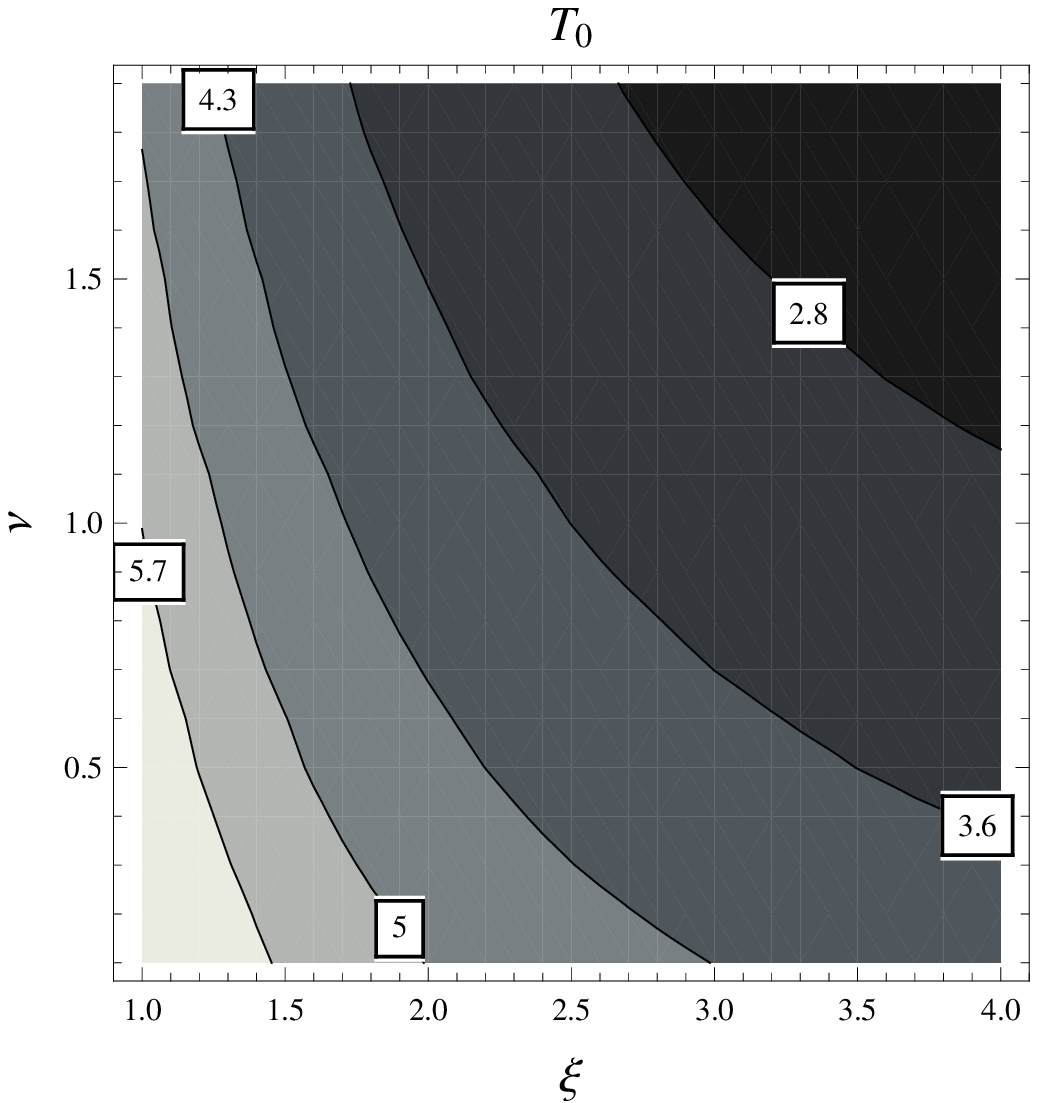}} \resizebox{65mm}{!}{\includegraphics{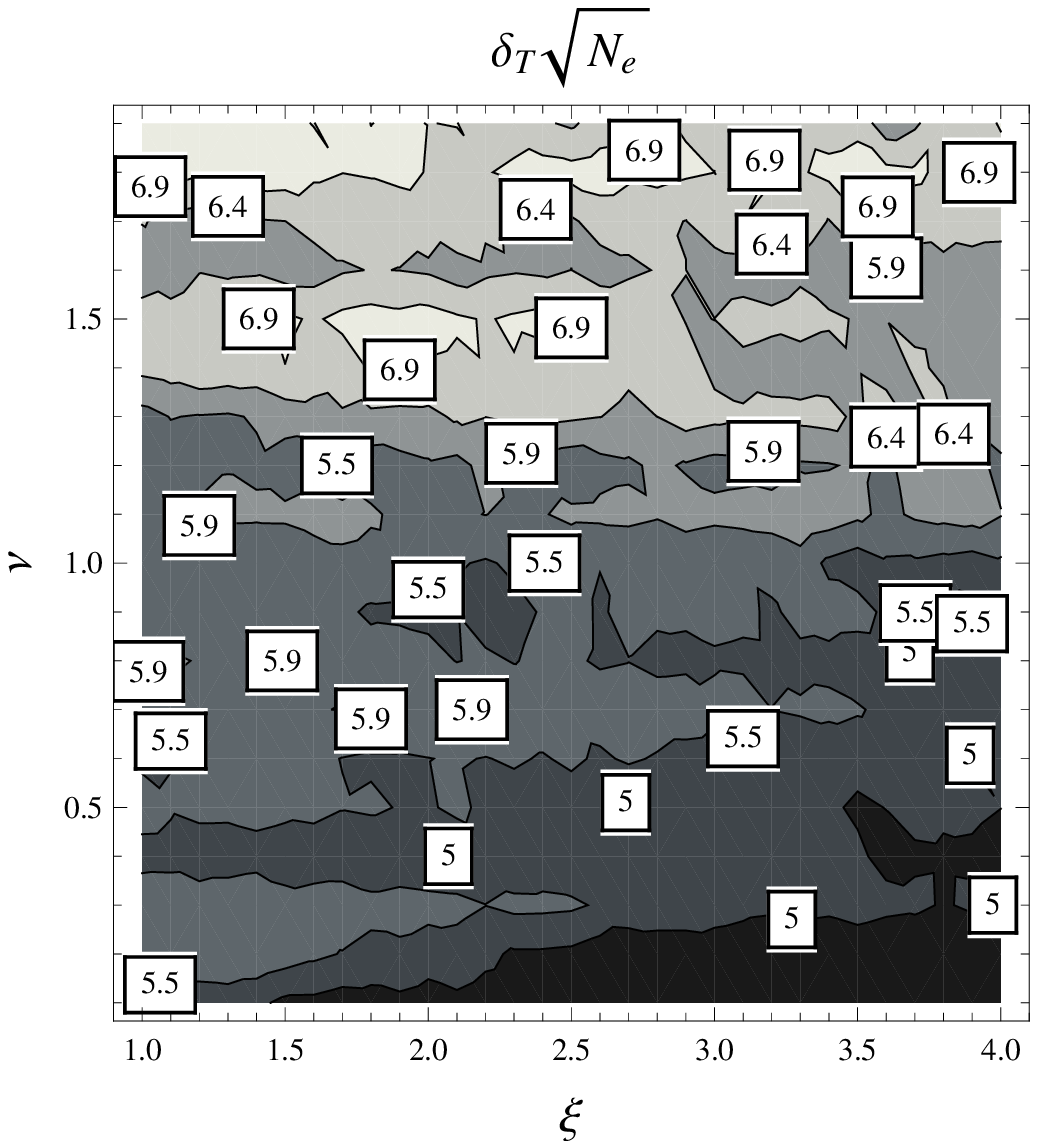}}\\
\end{center}
\caption{Instability growth time  and its root-mean-square
deviation.}
\label{Fig.2}
\end{figure}

In the numerical analysis  of statistical fluctuations of Cherenkov superradiance in the presence of shot noise, instead of the number
$N_e$ of real electrons, we took the number $N=360\xi\ll N_e$ of
large electrons  with initial phases equal to
\begin{equation}
\theta_\alpha(0)=\frac{2\pi \alpha}{N}+\sqrt{\frac{12N}{N_e}}r_\alpha,\alpha=1..N,
\end{equation}
where $r_\alpha$ are random variables uniformly distributed over the
interval $[0;1)$. It has been shown in  \cite{Penman1992} that
this procedure, boosting the performance of the program, simulates
the shot noise correctly. We selected the following values of
the controlling parameters:
 $n=N_e/\xi=2.7\cdot10^4,1.08\cdot10^5$, $\nu=0$---$2$, and $\xi=1$---$4$. The numerical experiment with each
($N_e$,$\xi$,$\nu$) triple was repeated 100 times.

Figures 1 and 2 show the results of computation from which we can
draw two very important conclusions. First, the instability growth time  and the peak radiated power, which is proportional to the reduced conversion ratio, are weakly
dependent on the number of particles $N_e$. In accordance  with \cite{Ginzburg2002}, the peak power increases with the bunch length $\xi$.
Second, the root-mean square deviations of $P_0$ and $T_0$ are inversely propotional to the square root of the number of particles: $\delta_P\approx11/\sqrt{N_e}$ and $\delta_T\approx6/\sqrt{N_e}$. 

At present, the electron beams for generating Cherenkov superradiance are obtained at high-current accelerators with explosive emission cathodes. Due to explosive electron emission, charged particles leave the cathode in separate portions, called ectons. A typical current of each ecton is $I_e\sim10$~A. The total current $I$ produced by an accelerator is several kiloamperes. As a result,  in estimating the fluctuations we should use the number of ectons $\sim I/I_e$ instead of $N_e$. Let the total current $I$ be 2.6~kA, then we have $\delta_P=0.68$ and $\delta_T=0.37$.  Let us note that the root-mean-square deviations  are of the same order of magnitude as their averages.

\section{Energy spread}

To take  account of the electron energy spread, we assume  the initial quantities $\dot\theta_\alpha(0)$ to be
Gaussian random variables whose averages equal zero and the
root-mean-square deviations $\sigma=\frac{C\Delta\gamma_\alpha}{\gamma_0^3}\sqrt[3]{\frac{v_0}{v_{gr}}}$ ($\Delta\gamma_\alpha$
is the root-mean-square deviation of the Lorenz factor).

\begin{figure}[ht]
\begin{center}
 \resizebox{65mm}{!}{\includegraphics{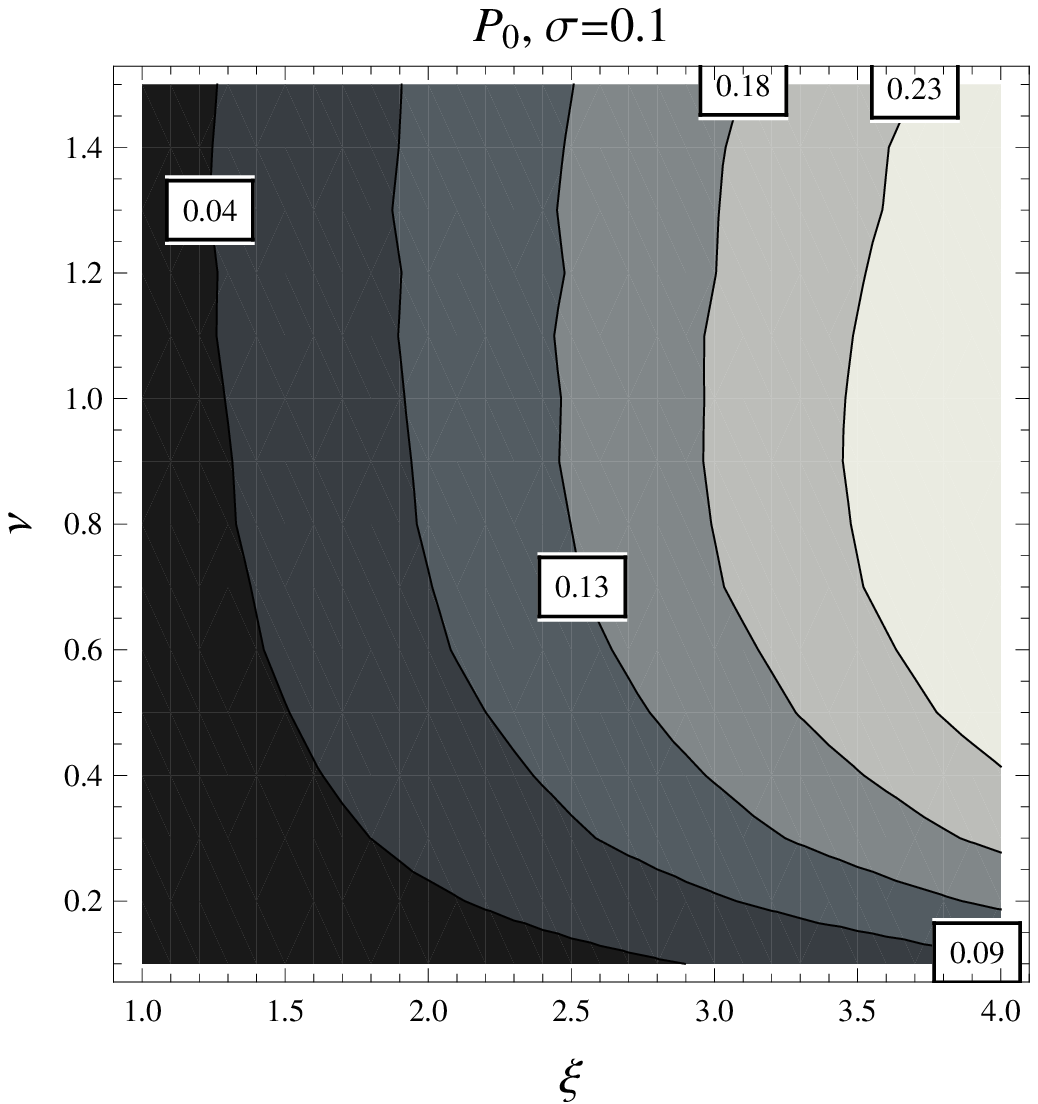}} \resizebox{65mm}{!}{\includegraphics{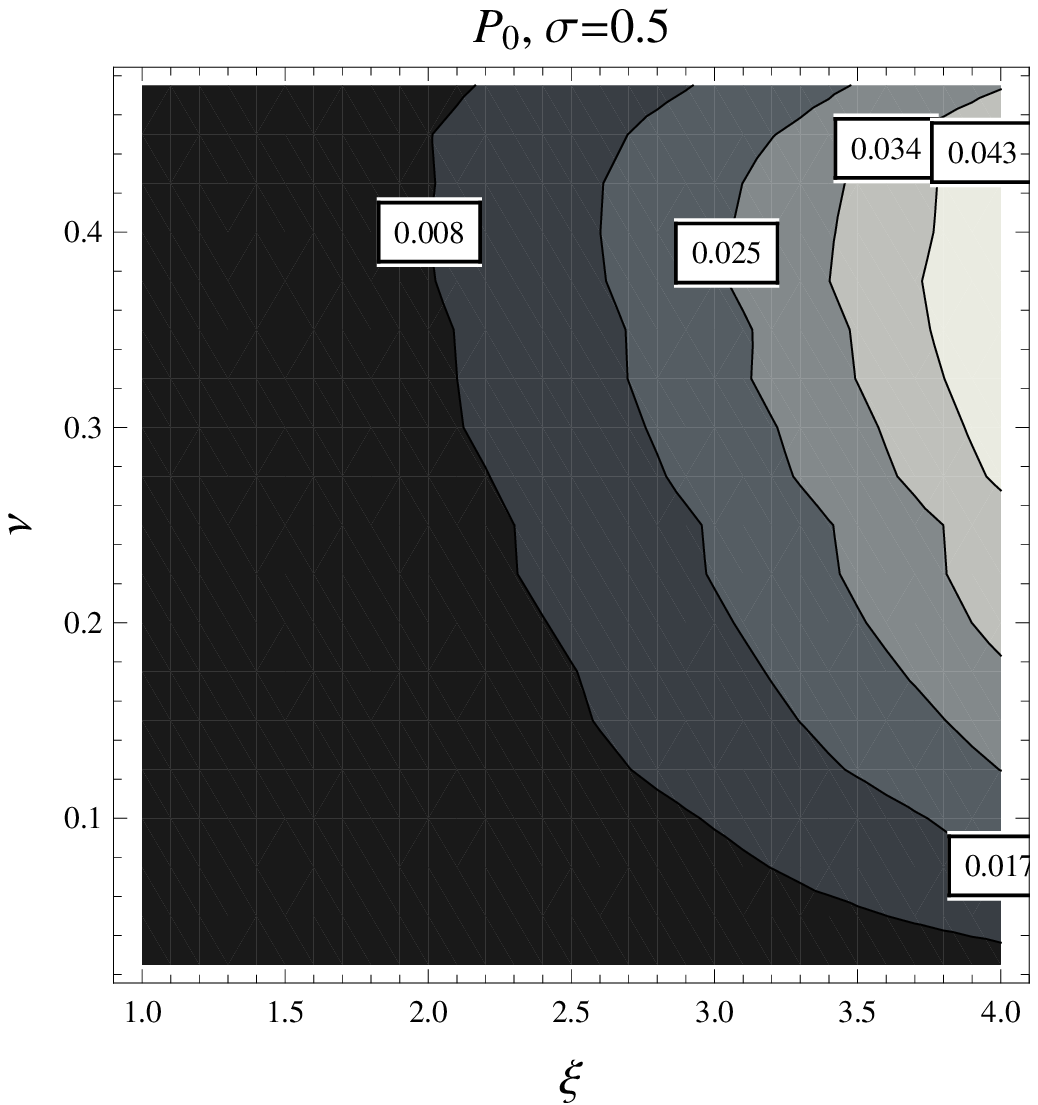}}\\
\end{center}
\caption{Reduced conversion ratio.} \label{Fig.3}
\end{figure}

Analyzing the results of numerical experiments, we can see that
the energy spread leads to a sharp
drop in the radiated power (Fig. 3), especially at high currents ($\nu\sim I_{av}$).
This is well-illustrated by the example in Fig. 3, where the growing influence of the energy spread with larger $\nu\sim I_{av}$ is seen clearly: the energy spread leads to a stronger suppression of radiation at large $\nu$.

\section{Quasi-Cherenkov superradiance}
This section considers quasi-Cherenkov superradiance emitted by electron bunches when charged particles pass through crystals  (natural or artificial) under the conditions of dynamical diffraction of electromagnetic waves.
We restrict ourselves to the case when the roots of the dispersion equation describing interaction between particles and electromagnetic waves don't coincide. The latter case demands more detailed computer simulations. This will be done in subsequent papers.

In the periodic structures, along with the electromagnetic wave emitted in the forward direction $F_0$, one can observe the electromagnetic wave that is emitted by charged particles in the diffraction direction $F_\tau$ and leaves the crystal through the bunch entrance surface.
The set of equations analogues to \eqref{bwo11} should be rewritten as follows:
\begin{equation}
\label{psi1}
\begin{split}
&\frac{\partial F_0}{\partial \tau}+\frac{\partial F_0}{\partial z}+i\chi F_\tau=-\frac{2}{N_\lambda}\sum_\alpha e^{-i\theta_\alpha},\\
&\frac{\partial F_\tau}{\partial \tau}-\frac{\partial F_\tau}{\partial z}+i\chi F_0=0,\\
&\frac{d^2\theta_\alpha}{d\tau^2}=-\Big(1+\nu\frac{d\theta_\alpha}{d\tau}\Big)^{3/2}\text{Re}(F_0e^{i\theta_\alpha}).\\
\end{split}
\end{equation} 
Here, $z$ and $\xi$ are defined as $z=C\beta x\sqrt[3]{v_0^2/v_{gr}^2}$ and $\xi=C\beta L\sqrt[3]{v_0^2/v_{gr}^2}$ instead of \eqref{bwo10}. The parameter $\chi$ in \eqref{psi1} is proportional to the dielectric susceptibility $\chi_\tau$ \cite{AnishchenkoBaryshevsky2015}. 

The equations \eqref{psi1} should be supplemented with boundary and initial conditions
\begin{equation}
\label{psi2}
\begin{split}
&\dot\theta_\alpha=0,\\
&F_0(0,\tau)=0,\\
&F_\tau(\Lambda,\tau)=0,\\
&\theta_\alpha=2\pi r_\alpha,
\end{split}
\end{equation} 
where $\Lambda$ is a crystal thickness and $r_\alpha$ are random variables uniformly distributed in the interval $[0;1)$.

In dimensionless units, the reduced conversion ratios are given by the expressions
\begin{equation}
\label{psi3}
\begin{split}
&P_0=\frac{\nu|F_{0peak}|^2}{8}\Big|_{z=\Lambda},\\
&P_\tau=\frac{\nu|F_{\tau peak}|^2}{8}\Big|_{z=0}.\\
\end{split}
\end{equation} 

We shall assume that $\nu=1.0$ and $=\xi=1.0$. For this case, the peak intensity of cooperative radiation emitted in forward and backward directions is investigated as a function of the crystal thickness $\Lambda$. The peak radiation intensity $P_0$ appeared to increase monotonically until saturation is achieved (Fig. 4). At saturation, fluctuations in the intensity of radiation undergo a sharp drop. 
The growth of parameter $\chi$ results in decreasing $P_0$ and increasing $P_\tau$ (Fig. 5, 6).
We would like to note the fluctuations of quasi-Cherenkov superradiance under dynamical diffraction conditions correlate well with the results obtained in the previous section for Cherenkov radiation.

\begin{figure}[ht]
\begin{center}
 \resizebox{65mm}{!}{\includegraphics{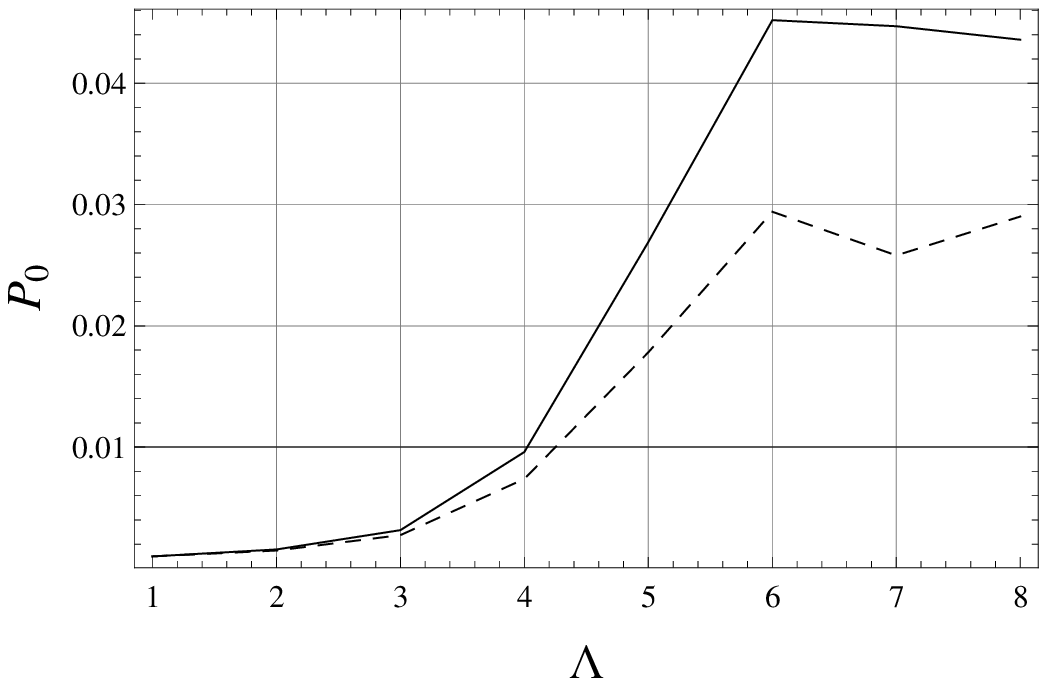}} \resizebox{65mm}{!}{\includegraphics{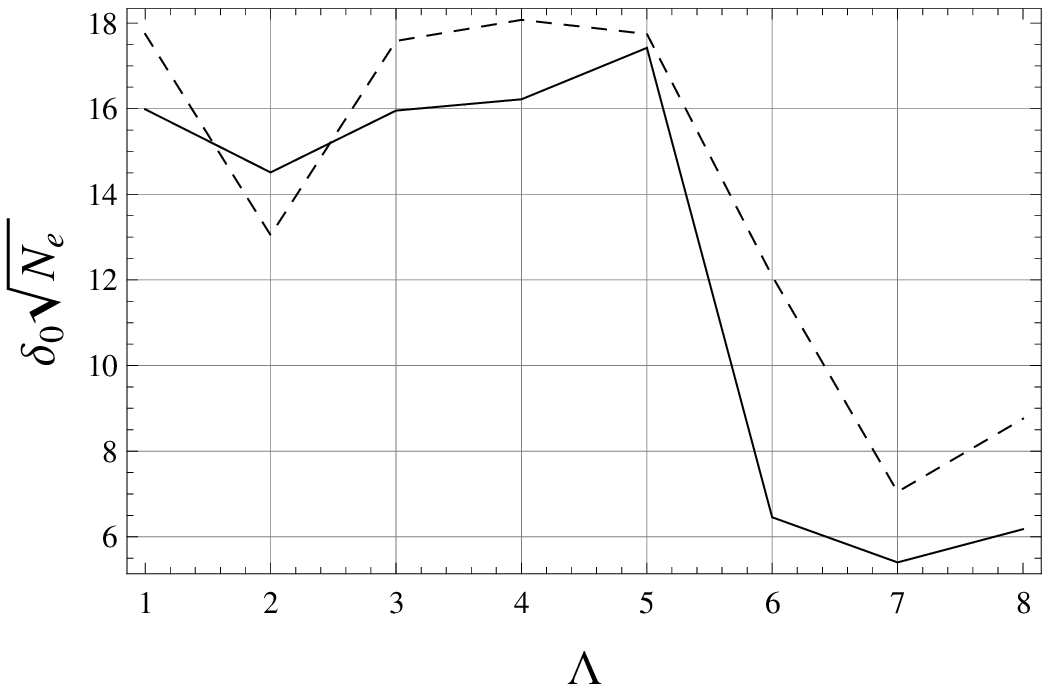}}\\
\end{center}
\caption{Quasi-Cherenkov radiation in forward direction [solid curve --- $\chi=0.1$, dashed curve --- $\chi=0.4$].} \label{Fig.4}
\end{figure}

\begin{figure}[ht]
\begin{center}
 \resizebox{65mm}{!}{\includegraphics{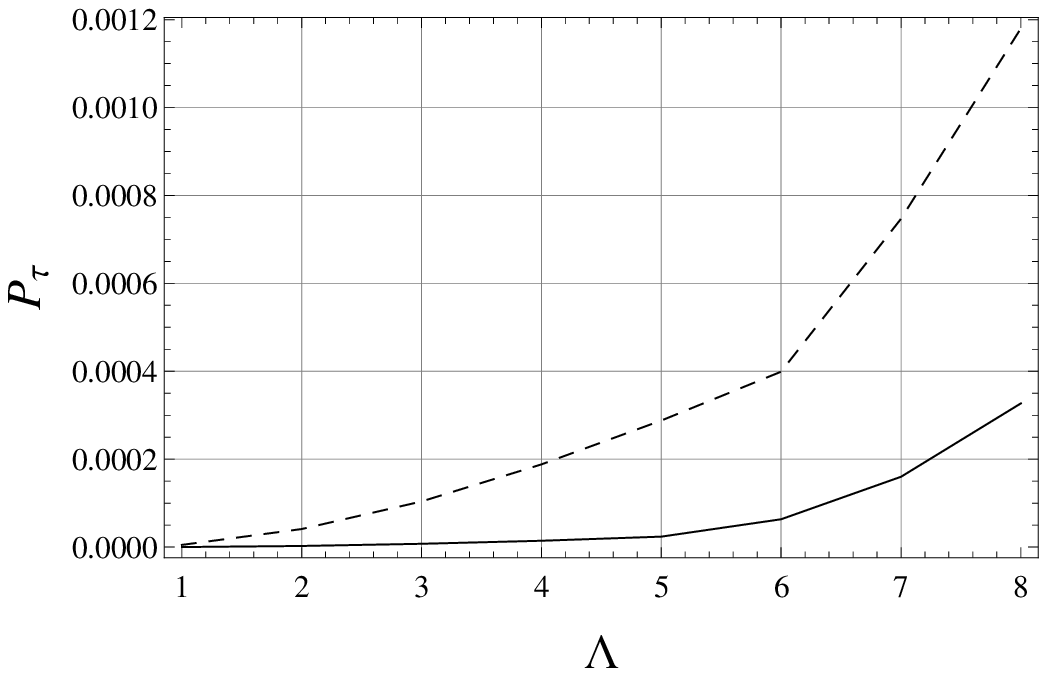}} \resizebox{65mm}{!}{\includegraphics{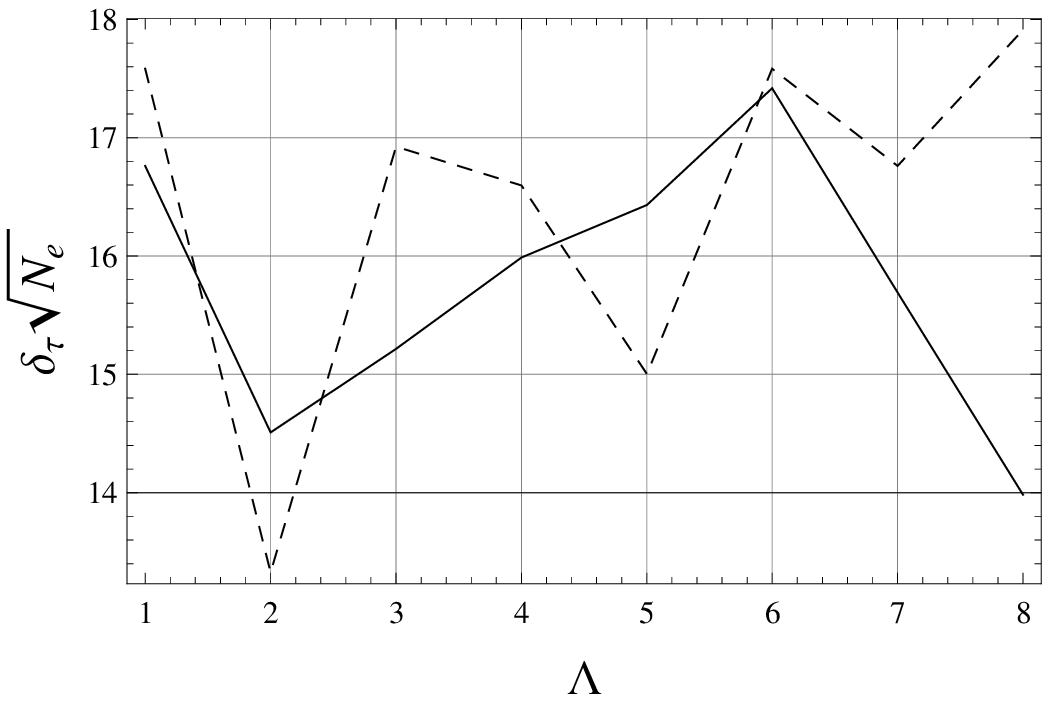}}\\
\end{center}
\caption{Quasi-Cherenkov radiation in backward direction [solid curve --- $\chi=0.1$, dashed curve --- $\chi=0.4$].} \label{Fig.5}
\end{figure}

\begin{figure}[ht]
\begin{center}
 \resizebox{65mm}{!}{\includegraphics{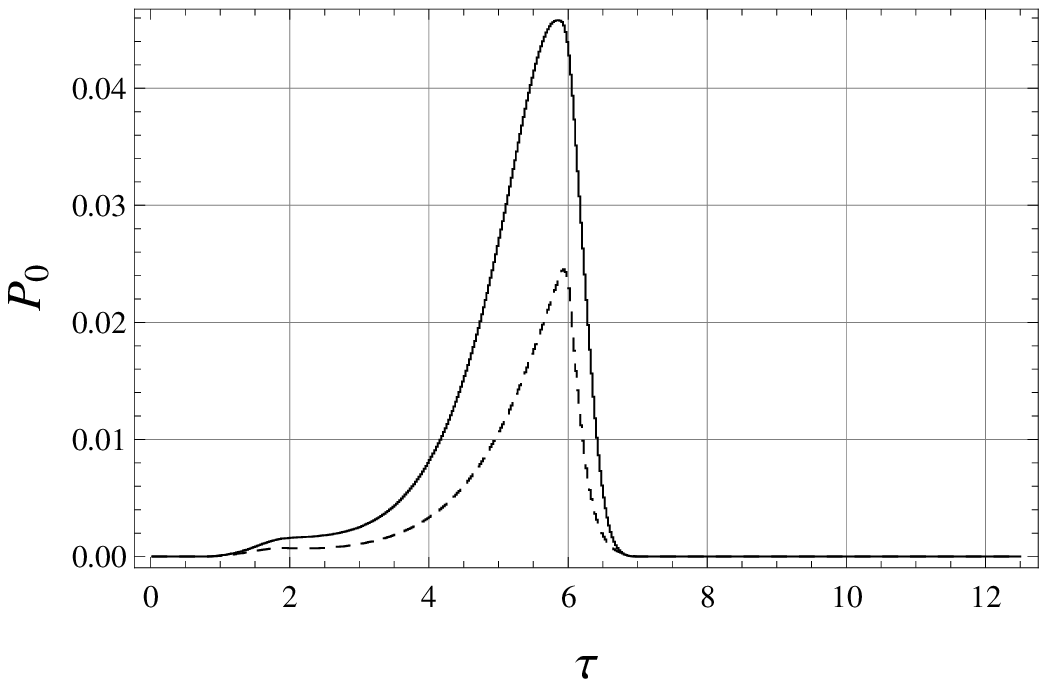}} \resizebox{65mm}{!}{\includegraphics{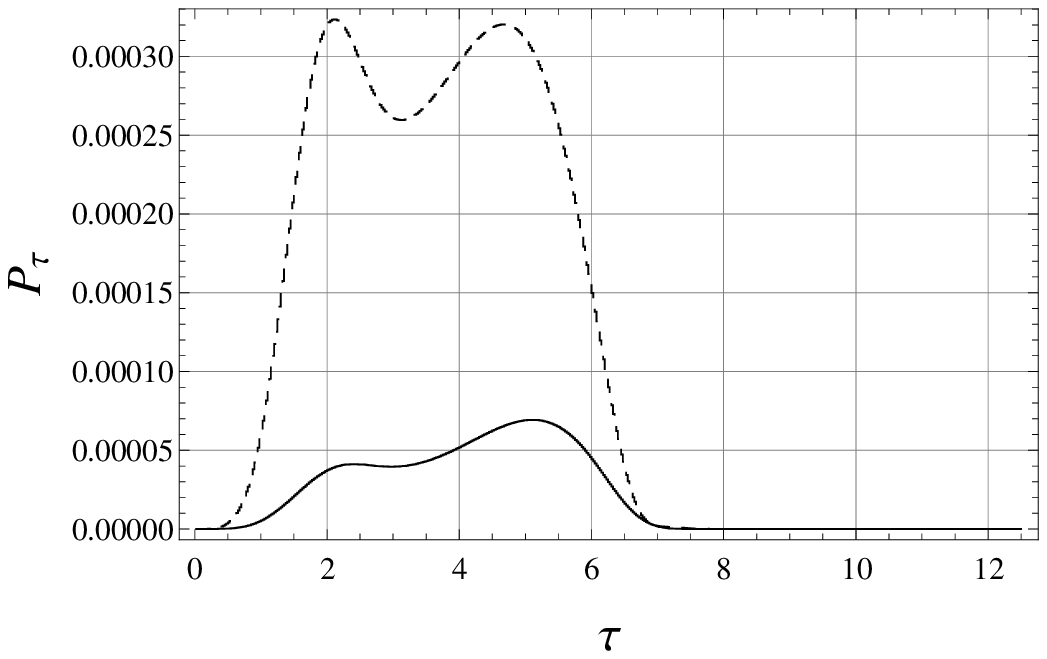}}\\
\end{center}
\caption{Quasi-Cherenkov radiation at saturation [solid curve --- $\chi=0.1$, dashed curve --- $\chi=0.4$].} \label{Fig.6}
\end{figure}

\section{Conclusion }
In this paper, we have studied the statistical properties of
Cherenkov and quasi-Cherenkov superradiance. For the Cherenkov superradiance, it has been shown that the relative root-mean-square
deviations of the radiated power and the instability growth time are
$\delta_P\approx11/\sqrt{N_e}$ and $\delta_T\approx6/\sqrt{N_e}$, respectively.
The fluctuations of quasi-Cherenkov superradiance under dynamical diffraction conditions correlate well with the results obtained for the Cherenkov superradiance.
While investigeting the quasi-Cherenkov superradiance, we have restricted ourselves to the case when the roots of the dispersion equation don't coincide.
The latter case demands more detailed computer simulations. This will be done in subsequent papers.

At present, electron beams for generating Cherenkov superradiance are obtained in high-current accelerators through explosive electron emission. As a result, the electron flow is emitted in separate portions, called ectons. To estimate $\delta_{P,T}$, we should use the number of ectons $\sim I/I_e\sim100\div1000$ instead of $N_e$. As a result, the root-mean-square deviations  in peak radiated power and instability growth time are comparable to their averages.

The particle energy spread  leads to a sharp decrease of the peak
radiated power. The influence of the energy spread grows with growing electron current.

\end{document}